\documentclass{jfm}

\usepackage{psfrag}
\usepackage{amsmath}
\usepackage{upmath}
\usepackage{amssymb}
\usepackage{amsbsy}
\usepackage{graphicx}
\usepackage{natbib}
\usepackage{epsfig}
\usepackage{lineno}
\usepackage[usenames]{color}

\newcommand{\com}[1]{\textcolor{black}{#1}}

\title[On the flow separation mechanism in the inverse Leidenfrost regime]{On the flow separation mechanism in the inverse Leidenfrost regime}

\author[J. Arrieta, A. Sevilla]{J. Arrieta$^{1}$ and A. Sevilla$^2$}

\affiliation{$^1$ Instituto Mediterr\'aneo de Estudios Avanzados, UIB-CSIC, 07190, Esporles, Baleares, Spain$^2$ Grupo de Mec\'anica de Fluidos, Departamento de Ingenier\'ia T\'ermica y de Fluidos, Universidad Carlos III de Madrid. Avda. de la Universidad 30, 28911, Legan\'es, Madrid, Spain\\[\affilskip]}

\begin{document}

\maketitle

\begin{abstract}
The inverse Leidenfrost regime occurs when a heated object in relative motion with a liquid is surrounded by a stable vapour layer, drastically reducing the hydrodynamic drag at large Reynolds numbers due to a delayed separation of the flow. To elucidate the physical mechanisms that control separation, here we report a numerical study of the boundary-layer equations describing the liquid-vapour flow around a solid sphere whose surface temperature is above the Leidenfrost point. Our analysis reveals that the dynamics of the thin layer of vaporised liquid controls the downstream evolution of the flow, which cannot be properly described substituting the vapour layer by an effective slip length. In particular, the dominant mechanism responsible for the separation of the flow is the onset of vapour recirculation caused by the adverse pressure gradient in the rearward half of the sphere, leading to an explosive growth of the vapour-layer thickness due to the accumulation of vapour mass. Buoyancy forces are shown to have an important effect on the onset of recirculation, and thus on the separation angle. Our results compare favourably with previous experiments.
\end{abstract}

\section{Introduction\label{sec:introduction}}

The reduction of hydrodynamic drag in liquid flows at macroscopic and microscopic scales is of paramount importance to improve the efficiency of many applications involving liquid-solid contact. A promising drag-reduction method that has been extensively studied is the generation of a thin layer of gas or a cloud of bubbles in the wall region to reduce the shear stress and, in the case of flows around bluff bodies, to delay flow separation, thereby reducing the pressure drag. The lubricating gas layer can be produced by different mechanisms, including surface superhydrophobicity~\citep{Rothstein2010}, microbubble injection and supercavitation~\citep{Ceccio2010} or surface heating~\citep{Bradfield1962,Vakarelski2011,Vakarelski2014,Vakarelski2016,Vakarelski2017}, to name a few.

Heating a solid object immersed in a liquid to temperatures above the so-called \emph{Leidenfrost temperature} can lead to the generation of a stable vapour layer surrounding its surface, leading to the film-boiling heat transfer regime. This regime has been the subject of many investigations, partly due to its relevance during the quenching process in nuclear reactors, that is known to be relevant in nuclear accidents. An extensive review of the most relevant experimental and theoretical aspects of film-boiling heat transfer can be found in~\citet{Dhir1998}. Most of the previous theoretical studies were aimed at obtaining accurate predictions for the heat transfer coefficient, rather than focusing on the mechanisms associated with hydrodynamic drag reduction. One of the few exceptions is the work of~\citet{Bradfield1962}, who studied the flow around a heated hemisphere-cylinder model and reported an important drag reduction effect due to the formation of a stable vapour layer around the solid surface. The first measurement of drag of heated spheres above the Leidenfrost temperature in water was provided by~\citet{Zvirin1990}, who reported a reduction of less than $10\%$. Furthermore, for ambient liquid temperatures close to the saturation temperature,~\citet{Zvirin1990} observed a slender vapour cavity departing from the equator of the sphere. Later on, \citet{Liu1996} performed an extensive characterization of the film-boiling heat transfer of spheres immersed in water in natural- and forced-convection regimes. These authors noticed that, under certain conditions, the liquid-vapour flow remained attached along the whole surface of the sphere. On the theoretical side,~\citet{Wilson1979}, \citet{EpsteinHauser} and \citet{Bang1994} pioneered the study of the forced-convection film boiling regime, but none of these authors studied the flow around a sphere, nor the effect of the relevant control parameters on the resulting liquid-vapour flow.

More recently, the film-boiling regime has been extensively studied to explore its potential application as a method to reduce the drag of bluff bodies~\citep{Vakarelski2011,Vakarelski2014,Vakarelski2016,Vakarelski2017}. In the latter works, the drag reduction effect observed under film-boiling conditions was referred to as the \emph{inverse Leidenfrost phenomenon}. In particular,~\citet{Vakarelski2011} performed experiments with freely falling heated spheres in a quiescent liquid (FC-72 perfluorohexane), that reached near-terminal velocities in the large-Reynolds-number regime. The stable vapour layer that developed around the sphere was shown to reduce the hydrodynamic drag by over $85\%$. This drag reduction effect crucially depends on the separation angle which, in the aforementioned experiments, was found to be delayed up to $130^{\circ}$, in contrast with the much smaller value of about $80^{\circ}$ observed in the case without film boiling for values of the Reynolds below the drag crisis~\citep{Achenbach1972}. Later on,~\cite{Vakarelski2014} performed similar experiments using water as working fluid. In this case, a drag reduction of more than $75\%$ was also observed, but only in cases where the temperature of the ambient liquid was close to the saturation temperature. A detailed characterisation of the inverse Leidenfrost phenomenon was carried out by~\cite{Vakarelski2016} for different liquids having a wide range of viscosities, observing drag reduction effects consistent with the previous studies. Interestingly, when the heated solid sphere is released from the surrounding air atmosphere into a water tank close to the saturation temperature, its impact on the free surface induces the formation of a slender cavity that remains attached to the sphere, inducing a giant drag reduction effect of up to $99\%$~\citep{Vakarelski2017sa}.

Despite the considerable research effort devoted to the study of the flow around bluff bodies in the inverse Leidenfrost regime, there are still fundamental questions that remain unanswered. Among these questions, maybe the most important one concerns the mechanisms that control the separation of the liquid-vapour flow around the bluff body. Indeed, although several explanations have been proposed to explain the effect of the vapour layer on the observed drag reduction, none of them are based on solid hydrodynamic grounds. For instance,~\citet{Vakarelski2011} suggested that the vapour layer effectively transforms the no-slip boundary condition at the solid wall into a stress-free boundary condition at the liquid-vapour interface. However, the latter explanation cannot be correct, since the film-boiling regime would then correspond to the flow around a spherical bubble studied by~\citet{Moore1963}, who found, however, that there is no separation of the flow except in a very small region close to the rear stagnation point. An alternative explanation was proposed by~\citet{Vakarelski2016} and~\citet{Berry2017}, who suggested that the presence of the vapour layer induces an effective slip length at the solid wall. However, in the present work we demonstrate that, although the vapour layer does indeed induce a certain slip velocity at the interface, the Navier slip approximation used in previous works is not justified, since it overlooks the nontrivial mechanics associated with the vapour layer. Here we show that an appropriate description of the vapour layer dynamics is essential to understand the resulting two-phase flow around the solid.

Therefore, the present study aims at providing novel theoretical and numerical insights \com{into} the origin of flow separation in the inverse Leidenfrost regime. The paper is organized as follows. In the next section the formulation of the problem is derived. Section \S\ref{sec:results} is devoted to present the mechanism that control the delay of the separation of the flow, including the buoyancy-free case and the effect of buoyancy to compare with previous experimental results. The final section presents the concluding remarks.


\section{Formulation\label{sec:formulation}}

\begin{figure}
\begin{center}
\includegraphics[width=0.6\textwidth]{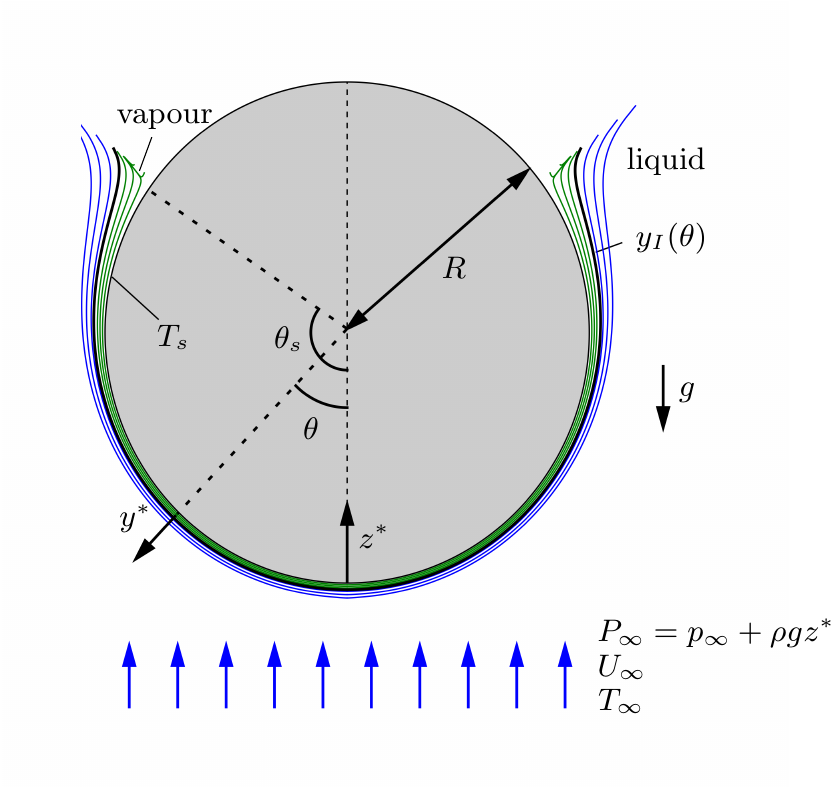}%
\caption{Sketch of the flow and definition of the main variables. Also shown are the interface position $y_I^*(\theta)$ (thick line), and several streamlines corresponding to the liquid and vapour boundary layers \com{(blue and green lines, respectively)}, for the case of water with $T_{\infty}=20^{\circ}$C, \com{$T_s=200^{\circ}$C}, in the buoyancy-free limit $Fr\to\infty$, obtained by numerically integrating equations~\eqref{cont_liquid_eq}--\eqref{energy_eq_interface}. Note that, for the vapour and liquid boundary layers to be observable at the scale of the sphere radius, $R$, the corresponding radial distances have been made dimensional assuming $Re=10^3$ and a sphere of radius $R=20$mm. The onset of vapour-flow reversal at \com{$\theta=\theta_s\simeq 125^{\circ}$} leads to a very fast increase of $y_I^*(\theta)$, that induces the appearance of a \emph{numerical} singularity at a certain angle \com{$\theta_f\simeq 127.2^{\circ} >\theta_s$}, at which the downstream marching scheme \com{employed in the present work fails} to converge. It is important to emphasise that the liquid boundary layer shows no sign of recirculation at $\theta_f$, indicating that the present separation phenomenon fundamentally differs from the classical one associated with a solid surface.}
\label{fig1}
\end{center}
\end{figure}

We consider the unbounded laminar axisymmetric flow around a sphere of radius $R$ with uniform surface temperature \com{$T_s$}. Far from the sphere, a stream of liquid of density \com{$\rho$, viscosity $\mu$}, temperature \com{$T_{\infty}$} and saturation temperature $T_{\mathrm{sat}}$, such that \com{$T_{\infty}<T_{\mathrm{sat}}<T_s$}, moves at constant velocity $U_\infty$ against the gravitational acceleration, $\mathbf{g}=-g\,\mathbf{e}_z$, being $\mathbf{e}_z$ the upwards direction, so that the modified pressure, \com{$p_{\infty}^*(z^*)+\rho g z^*$}, takes the constant value $P_{\infty}$ far from the sphere. The wall temperature, \com{$T_s$}, is assumed to be constant and above the Leidenfrost point~\citep{Quere2013}, so that the heat released from the sphere vaporizes the surrounding liquid and produces a thin vapour layer adjacent to the wall, as sketched in figure \ref{fig1}. Thus, the configuration under study here is equivalent to a solid sphere falling at the terminal velocity in the inverse Leidenfrost regime~\citep{Vakarelski2011,Quere2013,Vakarelski2014,Vakarelski2016}.

\subsection{Boundary layer approximation}

Before stating the mathematical model, we would like to point out that the physical properties of the working liquid and its associated vapour have been assumed to be independent of the temperature. Thus, in writing the conservation equations, the liquid and vapour densities, $\rho$ and $\rho_v$, dynamical viscosities, $\mu$ and $\mu_v$, thermal conductivities, $k$ and $k_v$, and specific heats at constant pressure, $c$ and $c_{pv}$, are all treated as constants that, in all the numerical results reported herein, were evaluated at the liquid film temperature, \com{$(T_{\infty}+T_{\mathrm{sat}})/2$}, and at the vapour film temperature, \com{$(T_s+T_{\mathrm{sat}})/2$}, respectively. \com{Another noteworthy aspect of the flow under study is the fact that the boundary layer remains laminar for values of the diameter-based Reynolds number substantially larger than the canonical value of $3\times 10^5$ of the single-phase incompressible boundary layer around a sphere~\citep{Berry2017}. Indeed, since the liquid tangential velocity at the liquid-vapour interface experiences only slight deviations from the potential velocity distribution in the region upstream of the separation point, the no-slip boundary condition that would apply to the liquid in the single-phase case is substituted here by a local effective slip condition that strongly stabilizes the liquid boundary layer, thereby justifying the assumption of laminar flow.}

\subsubsection{The boundary layer equations}

The Reynolds number, \com{$Re=\rho U_\infty R/\mu$, as well as the P\'eclet number, $Re Pr$, where $Pr=\mu c/k$ is the liquid Prandtl number, are both assumed to be large}. Thus, sufficiently upstream of the separation point, the liquid-vapour flow around the sphere can be described, in a first approximation, with use made of the boundary-layer form of the mass, momentum and energy conservation equations in spherical coordinates, together with far-field and wall boundary conditions, as well as appropriate matching conditions at the liquid-vapour interface, whose radial position $r^* = R + y_I^*(\theta)$ has to be obtained as part of the solution. As illustrated in figure~\ref{fig1}, the origin of the spherical coordinate system $(r^*,\theta,\varphi)$ is placed at the \com{centre} of the sphere, with the unit vector associated with the polar angle, $\theta$, pointing in the direction of the external stream at \com{$\theta=0^{\circ}$}. Note also that we have introduced the wall-normal coordinate $y^*= r^*-R$, appropriate for the boundary-layer formulation presented below. Here, starred variables are used to denote the dimensional variables that will have a dimensionless counterpart in the following development.

To facilitate the numerical integration, common characteristic scales were used to non-dimensionalise the conservation equations governing the liquid and the vapour phases, namely those associated with the \com{mechanical boundary layer of the liquid at the interface. Note that the relative thickness of the mechanical and thermal boundary layers is controlled by the liquid Prandtl number, which accomplishes the condition $Pr\gtrsim 1$ for the typical working liquids employed in the experiments. Consequently, the thickness of the velocity boundary layer is an appropriate length scale to non-dimensionalise the conservation equations. Specifically,} in terms of the wall-normal coordinate $y=y^*\sqrt{Re}/R$, the liquid and vapour polar velocities $(u,u_v)=(u^*,u_v^*)/U_\infty$, the corresponding radial velocities $(v,v_v)=(v^*,v_v^*)\sqrt{Re}/U_\infty$, the modified pressure $P=(p^*+\rho g z^*-P_{\infty})/(\rho U_{\infty}^2)$, and the reduced liquid and vapour temperatures, $\Theta=(T-T_\infty)/(T_{\mathrm{sat}}-T_\infty)$ and \com{$\Theta_v=(T_v-T_{\mathrm{sat}})/(T_{s}-T_{\mathrm{sat}})$}, the mass, polar momentum, and energy conservation equations for the liquid phase reduce to
\begin{align}
\frac{\p\,(u \sin{\theta})}{\p \theta}+\frac{\p\,(v \sin{\theta})}{\p y}=&0,
\label{cont_liquid_eq}\\
u\frac{\p u}{\p \theta}+v\frac{\p u}{\p y}=&
\frac{9}{8}\sin{2\theta} + \frac{\p^2 u}{\p y^2},
\label{azimth_mom_liquid_eq}\\
u\frac{\p \Theta}{\p \theta}+v\frac{\p \Theta}{\p y}=&\frac{1}{Pr}\,\frac{\p^2 \Theta}{\p y^2},
\label{energ_liquid_eq}
\end{align}
respectively, with the pressure gradient given by the outer potential flow, $-\frac{\p P}{\p\theta}=\frac{9}{8}\sin{2\theta}$. Equations~\eqref{cont_liquid_eq}--\eqref{energ_liquid_eq} must be integrated in the domain $y_I(\theta)\leq y<\infty$ with the boundary conditions $u-\frac{3}{2}\sin{\theta}=\Theta=0$ at $y\to\infty$. The corresponding non-dimensional equations for the vapour stream read
\begin{align}
\frac{\p\,(u_v \sin{\theta})}{\p \theta}+\frac{\p\,(v_v \sin{\theta})}{\p y}=& 0,
\label{cont_vapour_eq}\\
\frac{\rho_v}{\rho}\left(u_v\frac{\p u_v}{\p \theta}+v_v\frac{\p u_v}{\p y}\right)=& \frac{9}{8}\sin{2\theta} 
+ \left(1-\frac{\rho_v}{\rho}\right) \frac{\sin \theta}{Fr^2} + \frac{\mu_v}{\mu}\,\frac{\p^2 u_v}{\p y^2},
\label{azimth_mom_vapour_eq}\\
\frac{\rho_v}{\rho}\left(u_v\frac{\p \Theta_v}{\p \theta}+v_v\frac{\p \Theta_v}{\p y}\right)=&
\frac{1}{Pr_v}\,\frac{\mu_v}{\mu}\,\frac{\p^2 \Theta}{\p y^2},\\
\label{energ_vapour_eq}
\end{align}
to be integrated in the domain $0\leq y\leq y_I(\theta)$, with the wall boundary conditions $u_v=v_v=\Theta_v-1=0$ at $y=0$.
The matching conditions at the liquid-vapour interface are given by the mass, momentum and energy balances at $y=y_I(\theta)$, namely
\begin{align}
u-u_v=&0,\label{kinematic_cond_interface}\\
v-\frac{\rho_v}{\rho}\,v_v-\left(1-\frac{\rho_v}{\rho}\right)u\,\frac{{\rm d} y_I}{{\rm d}
\theta}=&0,\label{cont_eq_interface}\\
\frac{\p u}{\p y}-\frac{\mu_v}{\mu}\frac{\p u_v}{\p y}=&0\label{mom_eq_interface}\\
\Theta-1=\Theta_v=&0,\label{temp_eq_interface}\\
\frac{Ja}{Pr}\,\frac{\p \Theta}{\p y}-\frac{\mu_v}{\mu}\,\frac{Ja_v}{Pr_v}\,\frac{\p \Theta_v}{\p y}+\frac{\rho_v}{\rho}\left(v_v-u\frac{\mathrm{d} y_I}{\mathrm{d} \theta}\right)=&0,\label{energy_eq_interface}
\end{align}
representing the no-slip condition, the interfacial mass balance, the continuity of tangential stresses, the continuity of temperature and the energy balance between conductive heat fluxes and heat release due to vaporisation, respectively.

The dimensionless parameters governing the flow are, on the one hand, those related to the physical properties of the fluid, namely the liquid and vapour Prandtl numbers, $Pr=\mu c/k$ and $Pr_v=\mu_v c_{pv}/k_v$, and their corresponding density and viscosity ratios, $\rho_v/\rho$ and $\mu_v/\mu$, respectively. On the other hand, the flow is also controlled by the subcooling Jakob number, $Ja=c(T_{\mathrm{sat}}-T_{\infty})/L_v$ the superheat Jakob number, \com{$Ja_v=c_{pv}(T_{s}-T_{\mathrm{sat}})/L_v$}, where $L_v$ represents the heat of vaporisation of the fluid, and the Froude number, $Fr=U_{\infty}/\sqrt{g R}$. It is worth pointing out that, as usual in the boundary-layer approximation, the resulting set of equations~\eqref{cont_liquid_eq}--\eqref{energy_eq_interface} is independent of the Reynolds number, that only acts as a scaling factor in the wall-normal direction. In particular, our boundary-layer model predicts that the separation angle should not depend on the Reynolds number, provided that the interaction of the separated flow with the boundary layer is weak.

\subsubsection{Buoyancy forces and saturation temperature}

Note that we have neglected buoyancy forces in the liquid momentum equation~\eqref{azimth_mom_liquid_eq}, since the Archimedes number, $Ar=Rg\beta(T_{\mathrm{sat}}-T_\infty)/U_\infty^2=Gr/Re^2\ll 1$, where $\beta$ is the coefficient of thermal expansion of the liquid, and $Gr=R^3 g \beta (T_{\mathrm{sat}}-T_\infty)/\nu^2$ is the liquid Grashof number. Indeed, in the particular case of a sphere with $R=10$ mm~\citep{Vakarelski2014}, and assuming that $T_{\infty}=20^{\,\circ}$C, the Grashof number takes the values $Gr\simeq 1.1 \times 10^6$. Since $Re \gtrsim 10^4$ in the examples considered in the present work, the condition $Ar \ll 1$ is always satisfied, and thus liquid buoyancy effects can be neglected in a first approximation. In contrast, the buoyancy force has been retained in the vapour momentum equation~\eqref{azimth_mom_vapour_eq}, an effect that will be shown to play an important role under most realistic circumstances.

It should also be noted that the temperature of saturation $T_{\mathrm{sat}}$ has been assumed to remain constant along the liquid-vapour interface. Indeed, $T_{\mathrm{sat}}$ is given as a function of pressure by the Clausius-Clapeyron equation, from which it is deduced that, whenever $P_{\infty}\gtrsim O(\rho U_{\infty}^2)$, as corresponds to all the cases considered herein, the relative change in $T_{\mathrm{sat}}$ along the interface is $\Delta T_{\mathrm{sat}}/T_{\mathrm{sat}} \sim O(R_v T_{\mathrm{sat}_\infty}/L_v)$, where $T_{\mathrm{sat}_\infty}$ represents the temperature of saturation at ambient pressure, and $R_v=R^0/W$ is the vapour constant, with $R^0$ representing the universal gas constant and $W$ the molecular weight of the vapour. In the particular case of water, and assuming $T_{\mathrm{sat}_\infty}=100^{\circ}$C, the parameter $R_v T_{\mathrm{sat}_\infty}/L_v$ takes the values $0.077$, and the variations of $T_{\mathrm{sat}}$ can be neglected in a first approximation.

\subsubsection{Validity of the boundary-layer approximation}

\com{Apart from the validity conditions $Re\gg 1$ and $RePr\gg 1$, that are common to all boundary-layer analyses, the problem at hand presents specific aspects that must be carefully considered to ensure the slenderness of the two-phase flow around the sphere, and the assumption of negligible transverse pressure gradients. Indeed, one possible limitation of the classical boundary-layer equations~\eqref{cont_liquid_eq}--\eqref{energ_vapour_eq} may arise from the radial pressure variations induced by the vapour intake into the inner layer due to the liquid vaporisation at the interface. The latter transverse pressure variations must be compared with the polar pressure gradients induced by the acceleration of the external liquid stream. Note that the radial pressure gradient is especially relevant in cases where the temperature of the liquid is close to the saturation temperature, since most of the thermal energy released by the sphere wall is employed to vaporise the surrounding liquid, leading to the largest vapour intakes. Hence, to assess the validity of the boundary-layer approximation, an order-of-magnitude analysis has been carried out to estimate the characteristic transverse pressure gradients in the vapour stream when the liquid temperature is close to its saturation value. At leading order, both the non-dimensional characteristic thickness of the vapour layer, $\delta_v^*/R$, and the ratio of the radial pressure gradient to the polar pressure gradient, $\Delta p^*_y/\Delta p^*_\theta$, can be estimated by combining the following facts: i) The balance of vaporization enthalpy and heat flux coming from the vapour in the interfacial energy equation~\eqref{energy_eq_interface}, which yields $k_v(T_s-T_{\mathrm{sat}})/\delta_v^* \sim \rho_v L_v v_{v,c}^*$, where $v_{v,c}^*$ is the characteristic radial velocity of the vapour stream. ii) The balance between the outer pressure gradient and the radial diffusion of vapour momentum in the polar momentum equation~\eqref{azimth_mom_vapour_eq}, namely $\rho U_{\infty}^2/R \sim \mu_v u_{v,c}^*/\delta_v^{*2}$, where $u_{v,c}^*$ is the characteristic polar velocity of the vapour stream. iii) The vapour continuity equation~\eqref{cont_vapour_eq}, providing $v_{v,c}^*/\delta_v^*\sim u_{v,c}^*/R$. iv) The balance of transverse pressure gradient and radial diffusion of radial momentum in the radial vapour momentum equation, providing $\Delta p^*_y/\delta_v^* \sim \mu_v v^{*}_v/\delta_v^{*2}$ . The combination of these three balances yields the estimations
\begin{equation}
\left(\frac{\Delta p^*_y}{\Delta p^*_\theta}\right)^{1/2} \sim \frac{\delta^*_v}{R} \sim \left[\frac{Ja_v}{Pr_v}\frac{\rho}{\rho_v}\left(\frac{\mu_v}{\mu}\right)^2\right]^{1/4}\frac{1}{\sqrt{Re}}.
\label{deltav}
\end{equation}
For the temperature range $300< T_s < 700^{\,\circ}$C employed by~\cite{Vakarelski2014}, the prefactor of $Re^{-1/2}$ in the estimation~\eqref{deltav} varies from 1.06 to 1.91, what ensures the validity of the boundary layer approximation at high Reynolds numbers.}

\subsubsection{Numerical method}

The parabolic free-boundary problem given by equations~\eqref{cont_liquid_eq}--\eqref{energy_eq_interface} was numerically integrated using a front-fixed method~\citep{Crank1984}. In short, equations~\eqref{cont_liquid_eq}--\eqref{energy_eq_interface} were rewritten in terms of the normalised variable $Y=y/y_I(\theta)$, and integrated making use of a second-order implicit finite-difference scheme adapted from that proposed by~\citet{Anderson1984}. The radial position of the liquid-vapour interface, $y_{I}(\theta)$, is obtained as part of the solution.

\subsection{Stagnation-point flow}

The initial conditions for the integration of \eqref{cont_liquid_eq}--\eqref{energy_eq_interface} were obtained using their self-similar form near the forward stagnation point, $\theta\ll 1$~\citep{EpsteinHauser}, according to the expansion $u=F(y)\,\theta + O(\theta^2)$, $u_v=F_v(y)\,\theta + O(\theta^2)$, $v=V(y)+O(\theta)$, $v_v=V_v(y)+O(\theta)$, $p=-\frac{9}{8}\theta^2+O(\theta^4)$, $y_I=Y_I+O(\theta^2)$, with the constant $Y_I$ obtained as part of the solution. Using the latter expansion, the leading-order conservation equations for the liquid phase reduce to
\begin{align}
V'+2F&=0,\label{eq1}\\
F^2+VF'&=\frac{9}{4}+F'',\\
Pr\,V\,\Theta'&=\Theta'',
\end{align}
with the boundary conditions $F-\frac{3}{2}=\Theta=0$ at $y\to\infty$, while the vapour flow is governed by the system
\begin{align}
    V'_v+2F_v&=0,\\
    \frac{\rho}{\rho_v}\left(F_v^2+V_v F_v'\right)&=\frac{9}{4}+\frac{1}{Fr^2}(1-\frac{\rho_v}{\rho})+\frac{\mu_v}{\mu}\,F''_v,\\
    Pr_v \frac{\rho}{\rho_v}\,V_v\,\Theta'_v=&\frac{\mu_v}{\mu}\,\Theta''_v,
\end{align}
subjected to the boundary conditions $F_v=V_v=\Theta_v-1=0$ at $y=0$. Finally, the matching conditions at $y=Y_I$ take the simplified leading-order form
\begin{align}
   F-F_v&=V-\frac{\rho_v}{\rho},\\
   V_v&=\frac{\mu}{\mu_v}\,F'+F'_v,\\
   \Theta-1&=0,\\
   \Theta_v&=0,\\
   \frac{Ja}{Pr}\,\Theta'+\frac{\rho_v}{\rho} V_v&=\frac{\mu_v}{\mu}\frac{Ja_v}{Pr_v}\,\Theta'_v.\label{eq2}
\end{align}
where primes denote derivatives with respect to $y$. To determine the solution of~\eqref{eq1}--\eqref{eq2}, use was made of the normalised variable $Y=y/y_I(\theta)$, and the same spatial discretisation employed for the downstream evolution problem was used to obtain the solution of the stagnation point flow.

\subsubsection{Characteristic vapour-layer thickness}

\begin{figure}
\begin{center}
\includegraphics[width=1.0\textwidth]{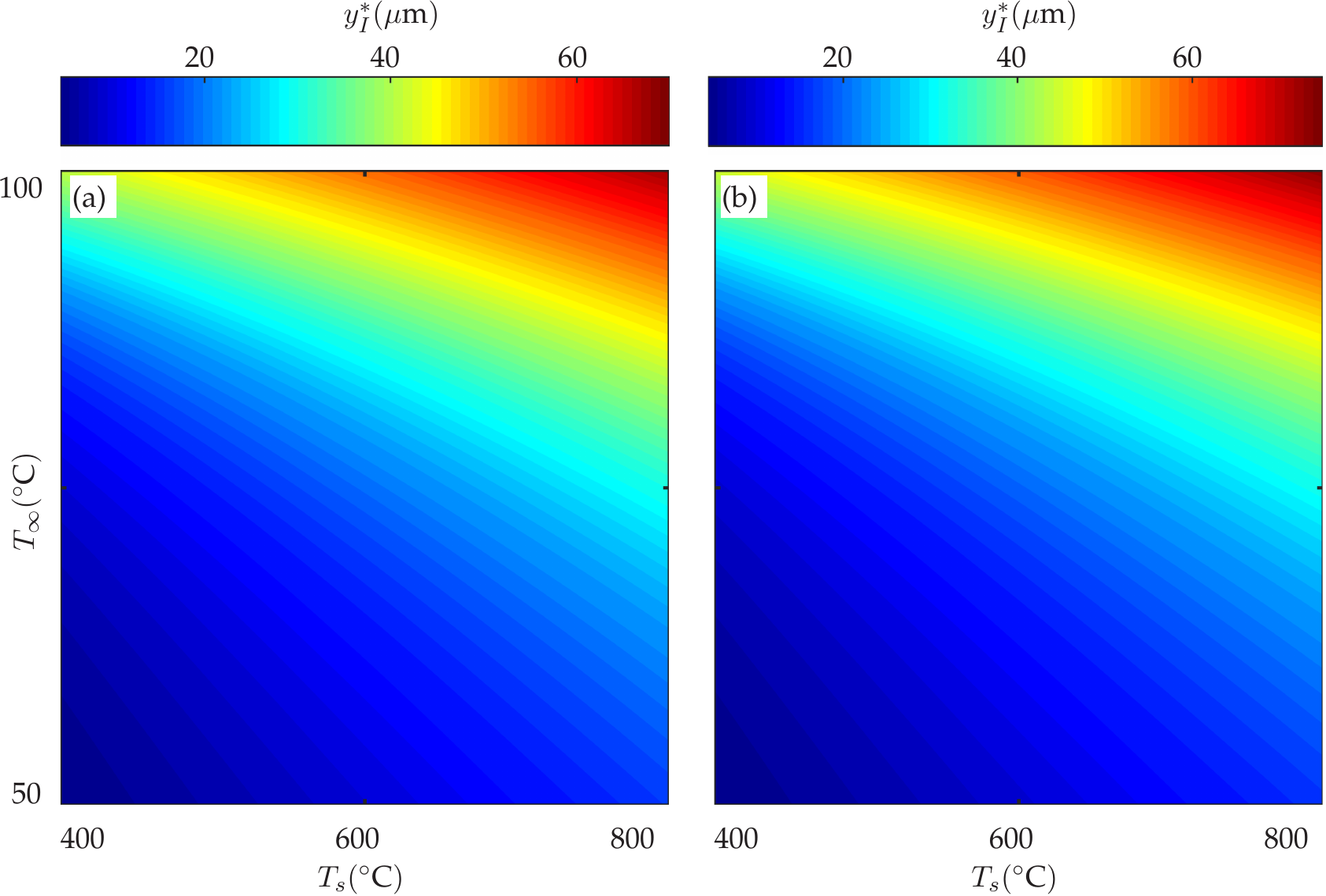}%
\caption{\com{The thickness of the vapour layer at the forward stagnation point, $y_I^*=Y_I R/\sqrt{Re}$ for $Re=10^5$ and $R=10$ mm~\citep{Vakarelski2014}. $Y_I$ was obtained by integrating the self-similar system of equations~\eqref{eq1}--\eqref{eq2} for (a) $Fr=1$, and (b) $Fr\rightarrow\infty$.}}
\label{figsweep}
\end{center}
\end{figure}

\com{The stagnation-point flow provides a useful way to estimate the vapour layer thickness, and to compare its characteristic value with those reported by~\cite{Vakarelski2014}. To that end, a parametric sweep of the self-similar problem~\eqref{eq1}--\eqref{eq2} was performed using water as working liquid, two values of the Froude number, and temperature ranges $T_s\in[400,800]^{\circ}$C and $T_\infty\in [50,100]^{\circ}$C. Figure~\ref{figsweep} shows contours of the dimensional vapour layer thickness, $y^*_I=Y_I R/Re^{1/2}$, assuming values of $Re=10^5$ and $R=10$ mm, which are typical values of the experiments reported by~\cite{Vakarelski2014}. The values of the Froude number are $Fr=1$ in figure~\ref{figsweep}(a), and $Fr\rightarrow\infty$ in figure~\ref{figsweep}(b). These plots reveal that the thickness of the vapour layer increases monotonically with the wall and free-stream temperatures. Moreover, it is deduced that buoyancy facilitates the downstream transport of vapour, resulting in slightly smaller thicknesses for the case with $Fr=1$ (figure~\ref{figsweep}a), compared with the case with $Fr\rightarrow\infty$ (figure~\ref{figsweep}b), with a maximum relative variation of about~$\lesssim 10\%$ between both cases. For the less sub-cooled cases, the vapour-layer thicknesses displayed in figure~\ref{figsweep} are in good agreement with the experiments of~\cite{Vakarelski2014} who, in the particular case of a quiescent sphere, reported values in the range $50\,\mu$m $\lesssim \delta_v^* \lesssim 150\,\mu$m. Indeed, note that the vapour layer is expected to be thicker for a quiescent sphere, since in the latter case the thickness results from a balance between buoyancy and viscous forces, whereas in the mixed-convection case considered herein the outer pressure gradient enhances the downstream transport of vapour, reducing the thickness of the vapour layer with respect to the natural convection case considered by~\cite{Vakarelski2014}.}

\section{Results\label{sec:results}}

\subsection{The buoyancy-free limit\label{subsec:fr_inf}}

The present section is devoted to study the buoyancy-free limit, $Fr\to\infty$, and to explain the flow separation mechanism. For a given working fluid, the only dimensionless parameters appearing in equations~\eqref{cont_liquid_eq}--\eqref{energy_eq_interface} are the subcooling and superheat Jakob numbers, $Ja$ and $Ja_v$, respectively. Note that water will be considered as the working fluid in the remainder of the manuscript.

\begin{figure}
\begin{center}
\includegraphics[width=1.0\textwidth]{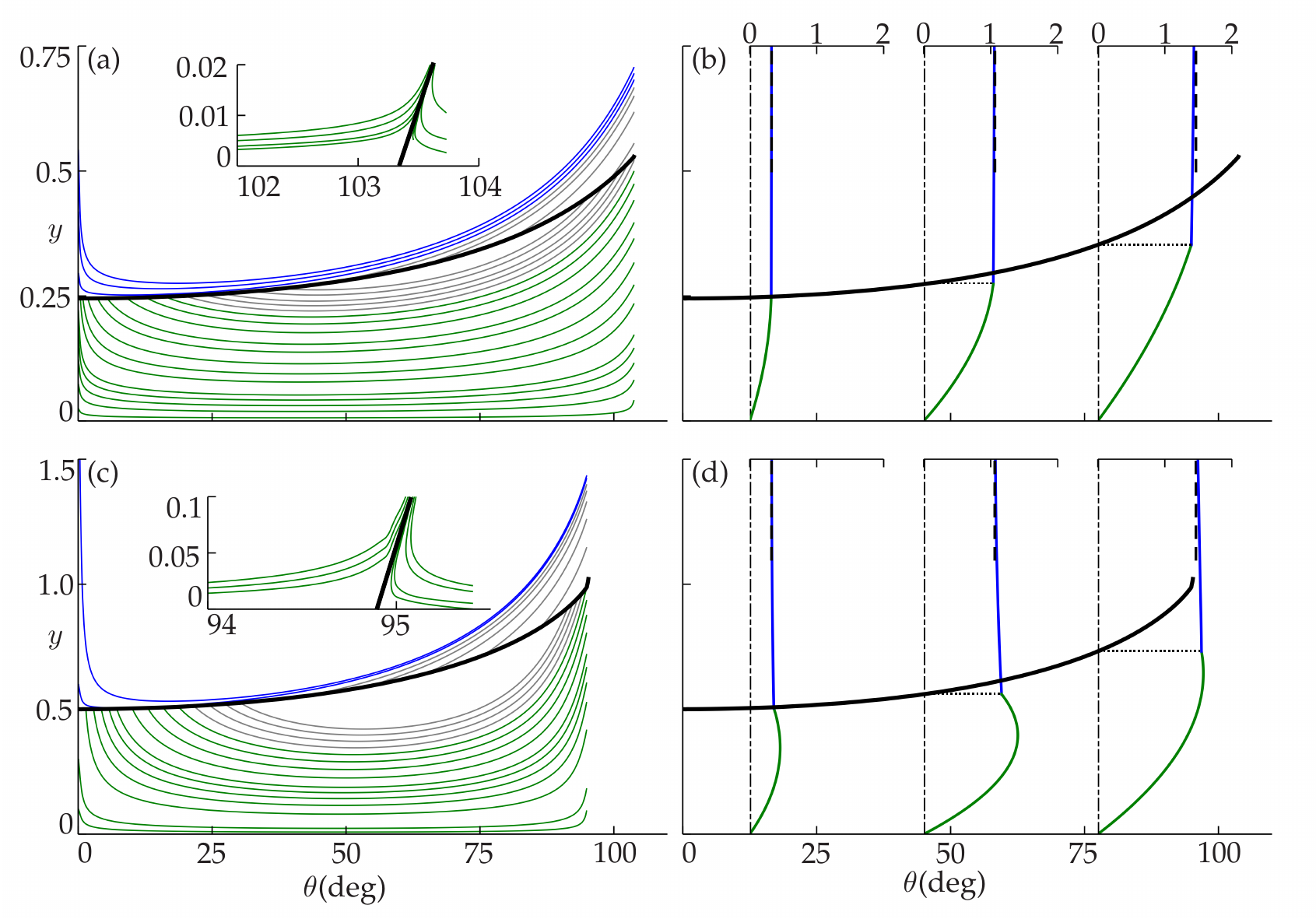}%
\caption{\com{Downstream evolution of the two-phase boundary layer in the buoyancy-free limit, $Fr\to\infty$. (a), (c) Streamlines of the liquid and vapour boundary layers (blue and green lines, respectively) for the cases (a) $T_\infty=50^{\circ}$C ($Ja=0.093$) and \com{$T_s=500^{\circ}$C} ($Ja_v=0.356$), and (c) $T_\infty=75^{\circ}$C ($Ja=0.046$) and \com{$T_s=500^{\circ}$C} ($Ja_v=0.356$). The effect of condensation is illustrated by gray streamlines the attach to, and depart from the interface in the vapour and liquid streams, respectively. (b), (d) Green solid lines represent radial profiles of vapour polar velocity, $u_v(y),\,0\leq y\leq y_I$, while $u(y),\,y\geq y_I$ is represented by blue solid lines at the downstream positions $\theta=12.5^{\circ}$, $\theta=45^{\circ}$ and $\theta=77.5^{\circ}$ for the same parameters as panels (a) and (c), respectively. The polar velocity of the outer flow at the aforementioned positions is represented with black dashed lines in panels (b) and (d). The onset of vapour recirculation takes place at \com{$\theta=\theta_s\simeq 103.4^{\circ}$} in panels (a) and (c), while \com{$\theta_s\simeq 95.1^{\circ}$} in panels (b) and (d). A detail of the onset of this recirculation is shown in the insets of panels (a) and (d), with the dividing streamline that separates the forward and backward flow plotted as a black solid line. In the four plots, the position of the interface, $y_I(\theta)$, is plotted with a thick black line.}\label{streamlines}}
\end{center}
\end{figure}

\subsubsection{Description of the flow evolution}

The downstream development of the flow is illustrated in figures~\ref{streamlines}(a) and (c), which show the interface position, $y_I(\theta)$ (thick solid black line), together with several vapour and liquid streamlines (\com{green and blue lines}, respectively), resulting from the integration of the system~\eqref{cont_liquid_eq}--\eqref{energy_eq_interface} with $T_\infty=50^{\,\circ} $C, and \com{$T_s=500^{\,\circ}$C} in figure~\ref{streamlines}(a), and $T_\infty=75^{\,\circ} $C, and \com{$T_s=500^{\,\circ}$C} in figure~\ref{streamlines}(c). In addition, figures~\ref{streamlines}(b) and (d) display the function $y_I(\theta)$ (thick solid black line), and several radial profiles of polar velocity (\com{green and blue lines for the vapour and the liquid, respectively}) at three different positions along the sphere indicated by vertical dash-dotted lines. These results reveal that the \com{thickness of the vapour layer} increases monotonically along the sphere until a certain angle $\theta_f$ is reached beyond which convergence cannot be achieved, at least with our numerical method. In particular, \com{$\theta_f=103.7^{\circ}$} for $T_\infty=50^{\circ}$C and \com{$\theta_f=95.7^{\circ}$} for $T_\infty=75^{\circ}$C. Notice from figures~\ref{streamlines}(b) and (d) that the liquid velocity profile is almost uniform along the entire sphere \com{and close to the value of the outer potential flow, represented by thick dashed lines at the corresponding station}, indicating that the \com{relative velocity increments in the liquid boundary layer due to the shear stress exerted by the vapour are small} in both cases. In contrast, the velocity of the vapour stream is strongly affected by the degree of subcooling. As the free-stream liquid temperature becomes closer to the saturation temperature, the mass of vapour produced is larger, increasing the thickness of the vapour layer. The increased vapour mass flux also affects the shape of the vapour velocity profiles, which become parabolic for the less subcooled case (figure~\ref{streamlines}d). Moreover, in this case the mean velocity of the vapour stream, and in particular the velocity at the liquid-vapour interface, are both larger than the liquid velocity outside the boundary layer.

\begin{figure}
\begin{center}
\includegraphics[width=1.0\textwidth]{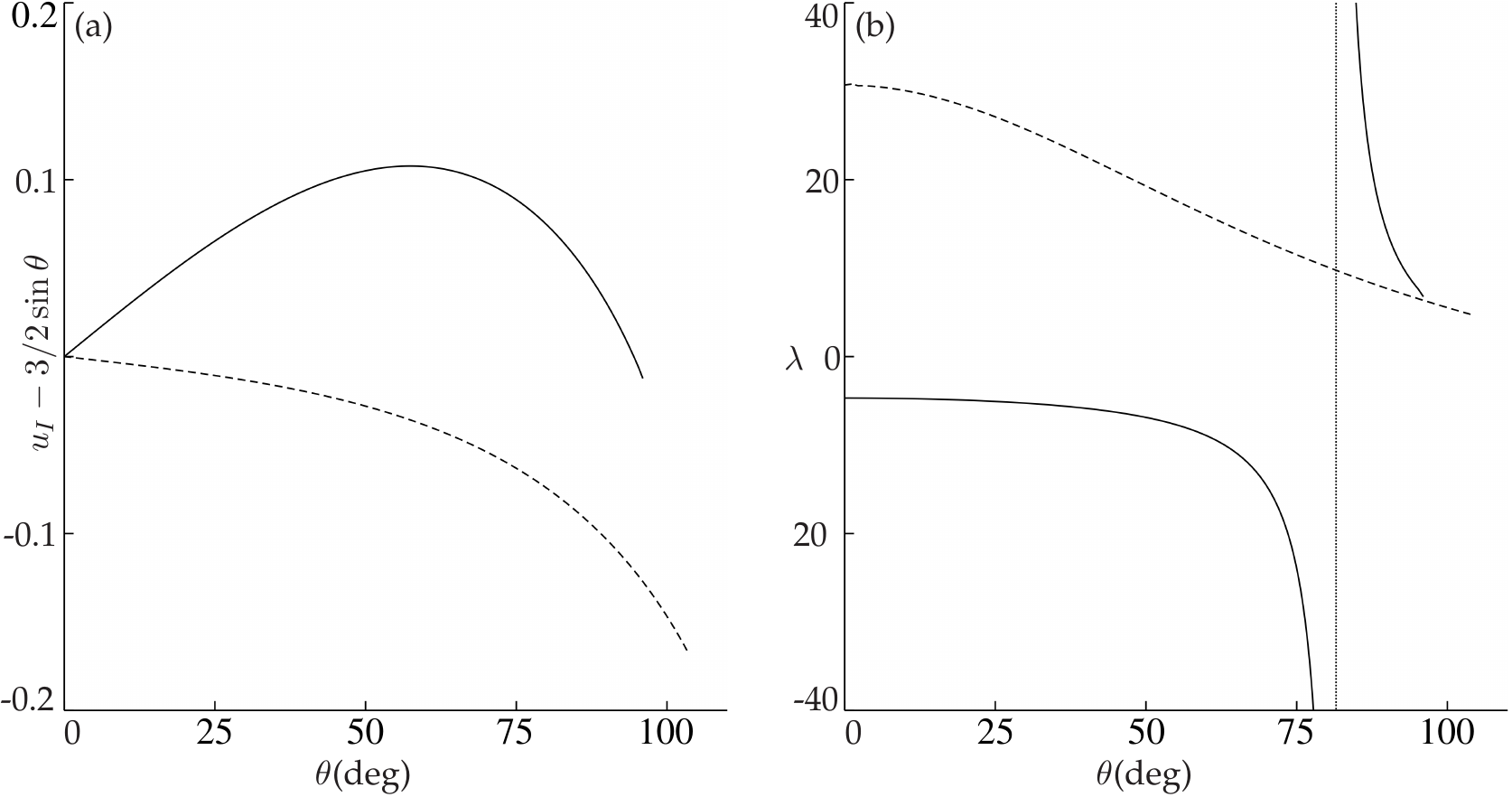}%
\caption{(a) The difference of the velocity at the interface, $u_I$ and the liquid velocity outside the boundary layer, $3/2\sin{\theta}$, for a wall temperature \com{$T_s=500^{\circ}$C}, and free-stream temperatures of $T_\infty=50^{\circ}$C (dashed line), and $T_\infty=75^{\circ}$C (solid line). (b) The slip length, $\lambda$, defined as the ratio of the velocity at the interface, $u_I$, and the slope of the velocity at the interface, $\p u/\p y|_{y_I}$, for \com{$T_s=500^{\circ}$C} and $T_\infty=50^{\circ}$C (dashed line) and $T_\infty=75^{\circ}$C (solid line). The location where the slip length becomes infinite is represented by the dotted vertical line.}
\label{fig3}
\end{center}
\end{figure}

The increasingly larger adverse pressure gradient for $\theta>90^{\circ}$ strongly decelerates the vapour stream, until an angle $\theta_s$ is reached where ${\p }u_v/{\p}y=0$, indicating the onset of a vapour recirculation bubble. In particular, vapour recirculation starts at \com{$\theta_s\simeq 103.4^{\circ}<\theta_f$}, for $T_\infty=50^{\circ}$C, and \com{$\theta_s\simeq 95.1^{\circ}$} for $T_\infty=75^{\circ}$C, as shown in the insets of figures~\ref{streamlines}(a) and figure~\ref{streamlines}(c). From these figures it is also deduced that the vapour layer thickness grows very fast just after the onset of recirculation, leading to the separation of the two-phase boundary layer. \com{In particular, the departing angles of the dividing streamlines represented by the black solid lines in the insets of figures~\ref{streamlines}(a) and (c) are 74$^{\circ}$ and 76$^{\circ}$, respectively}. The appearance of reverse flow in the vapour stream leads to numerical difficulties in the integration of the boundary-layer equations~\eqref{cont_liquid_eq}-\eqref{energ_vapour_eq}, which preclude the downstream marching and explain the numerical singularity encountered at $\theta_f$.

\subsubsection{The flow separation mechanism}

In view of the previous observations, the mechanisms that lead to the explosive growth of the vapour-layer thickness past the angle $\theta_s$ can be understood by taking into account the effect of the longitudinal pressure gradient, imposed by the outer potential flow on both the liquid and the vapour boundary layers, together with simple considerations based on the energy balance at the interface. Indeed, equation~\eqref{energy_eq_interface} indicates that the energy that arrives to the interface by conduction from the wall is employed to heat the liquid up to the saturation temperature, $T_{\mathrm{sat}}$, and the excess heat is responsible for the liquid-vapour phase change \com{as shown in figures~\ref{streamlines}(a) and (c) by the} streamlines that emerge from the interface due to liquid vaporisation, thus contributing to the injection of fresh fluid into the vapour boundary layer. Since the thickness of the vapour stream increases monotonically due to the accumulation of vaporised liquid, the heat flow towards the liquid continually decreases downstream. Eventually, a certain angle is reached where the energy supplied by the hot wall is only able to increase the temperature of the liquid up to $T_{\mathrm{sat}}$, and downstream from this point the energy required for the liquid to reach $T_{\mathrm{sat}}$ is supplied by vapour condensation, as illustrated \com{in figures~\ref{streamlines}(a) and (c) by the vapour streamlines that reattach to the interface close to $\theta_f$, and by the liquid streamlines that depart from the interface, both plotted in gray}. Although condensation removes vapour from the inner layer, and therefore tends to decrease the slope of the interface, the latter effect is counter-balanced by the adverse pressure gradient associated with the outer potential flow, together with the reduced area per unit streamwise length, both effects contributing to the fast increase of the interface slope past the angle $\theta=90^{\circ}$. Indeed, since both the density and the dynamic viscosity of the vapour are much smaller than the corresponding values for the liquid phase, the pressure gradient has a much stronger effect on the vapour flow than on the liquid flow, eventually leading to the onset of vapour recirculation at a certain angle $\theta_s$. Moreover, for $\theta>90^{\circ}$, the area per unit streawmwise length decreases due to the geometry of the sphere, thus hindering the downstream transport of the evaporated liquid. These two effects provide the explanation for the fast increase of the interfacial slope, that eventually leads to the separation phenomenon observed in the experiments~\citep{Vakarelski2011,Vakarelski2014,Vakarelski2016}.

It should be highlighted that the separation mechanism explained in the previous paragraph fundamentally differs from that proposed by~\citet{Vakarelski2011}, where the interface was assumed to behave as a shear-free layer due to the smallness of the vapour-to-liquid density and viscosity ratios. However, our numerical results show that the viscous shear stress does not vanish at the interface, as can be appreciated in figure~\ref{streamlines}(b) \com{and (d)}. Moreover, the growth of the recirculation bubble in the vapour stream forces the interface slope to increase very fast to enable the downstream transport of the accumulated vapour, whereas the liquid boundary-layer remains almost unaffected by the effect of the pressure gradient imposed by the outer potential flow and the growth of the vapour-layer thickness. In fact, the liquid flow shows no sign of recirculation near separation, in contrast with the classical separation scenarios associated with a solid wall~\citep{Schlichting} or a stress-free interface~\citep{Leal89,Blanco95}.

\subsubsection{Effective slip length}

Recent attempts to understand the observed drag reduction have studied the flow around a sphere replacing the no-slip boundary condition by an effective slip velocity, $u_I$, that depends on an arbitrarily defined slip length $\lambda=\lambda^*\sqrt{Re}/R$, and the radial gradient of the polar velocity at the interface~\citep{Vakarelski2016,Berry2017}
\begin{equation}
u_{I}=\lambda \left.\frac{\partial u}{\partial y}\right|_{y=y_I},
\label{eq:slip_length}
\end{equation}
where the value of $\lambda$ is arbitrarily chosen. This approach, that has been used in the study of the flow around superhydrophobic surfaces~\citep{McHale2011,Gruncell2013}, completely overlooks the dynamics of gas phase. However, here we demonstrate that, at least in the case of the inverse Leidenfrost regime, the dynamics of the vapour stream is essential, in that it controls the explosive growth of the interface leading to flow separation. In particular, the degree of liquid subcooling strongly affects the vaporisation rate. Indeed, the amount of vapour produced increases as the ambient liquid temperature approaches the saturation temperature, \com{since most of the thermal energy} coming from the wall is employed to vaporise. Consequently, the mean velocity of the gas stream, and in particular the velocity at the interface, may become larger than the velocity of the outer potential flow, as was already mentioned in \S\ref{subsec:fr_inf}. The latter effect can be appreciated in figure~\ref{fig3}, where the difference between the outer potential velocity and the interfacial velocities is represented for the same cases shown in figure~\ref{streamlines}. In the less subcooled case (solid line), the difference is negative, since the interface is accelerated by the large amount of vapour injected to the inner layer. This effect can also be appreciated in the velocity profiles plotted in figure~\ref{streamlines}(d) \com{at $\theta=45^{\circ}$ and $\theta=77.5^{\circ}$}, where it is seen that the slope of the liquid stream is slightly negative and \com{the velocity at the interface is slightly larger than the corresponding value of the outer flow, represented by the thick dashed lines.}

Figure~\ref{fig3}(b) shows the downstream evolution of the slip length as a function of the 
polar angle. The dashed line corresponds to the case with  \com{$T_s=500^{\,\circ}$C} and $T_\infty=50^{\,\circ}$C, whereas the dashed line was computed for \com{$T_s=500^{\,\circ}$C} and $T_\infty=75^{\,\circ}$C. 
Unsurprisingly, the value of the slip length is not constant along the sphere, since the 
dynamics of the vapour layer modifies both the slip velocity and the interfacial shear stress in a nontrivial way. In particular, the large vapour velocities associated with small degrees of subcooling impose a negative value to $\partial u/\partial y$ at the interface, thereby leading to negative values of the slip length. Furthermore, at the downstream position where the liquid slope becomes zero, the slip length becomes infinite, as can be seen in figure~\ref{fig3}(b) for the case \com{$T_s=500^{\,\circ}$C}, $T_\infty=75^{\,\circ}$C (solid line). The latter facts highlight the importance of studying the dynamics of the gas phase in two-phase drag reduction configurations, which cannot be properly described using a constant effective slip length\com{, as already pointed out by~\cite{Berry2017}.}

\subsection{The effect of buoyancy}\label{subsec:buoyancy}

\begin{figure}
\begin{center}
\includegraphics[width=1.0\textwidth]{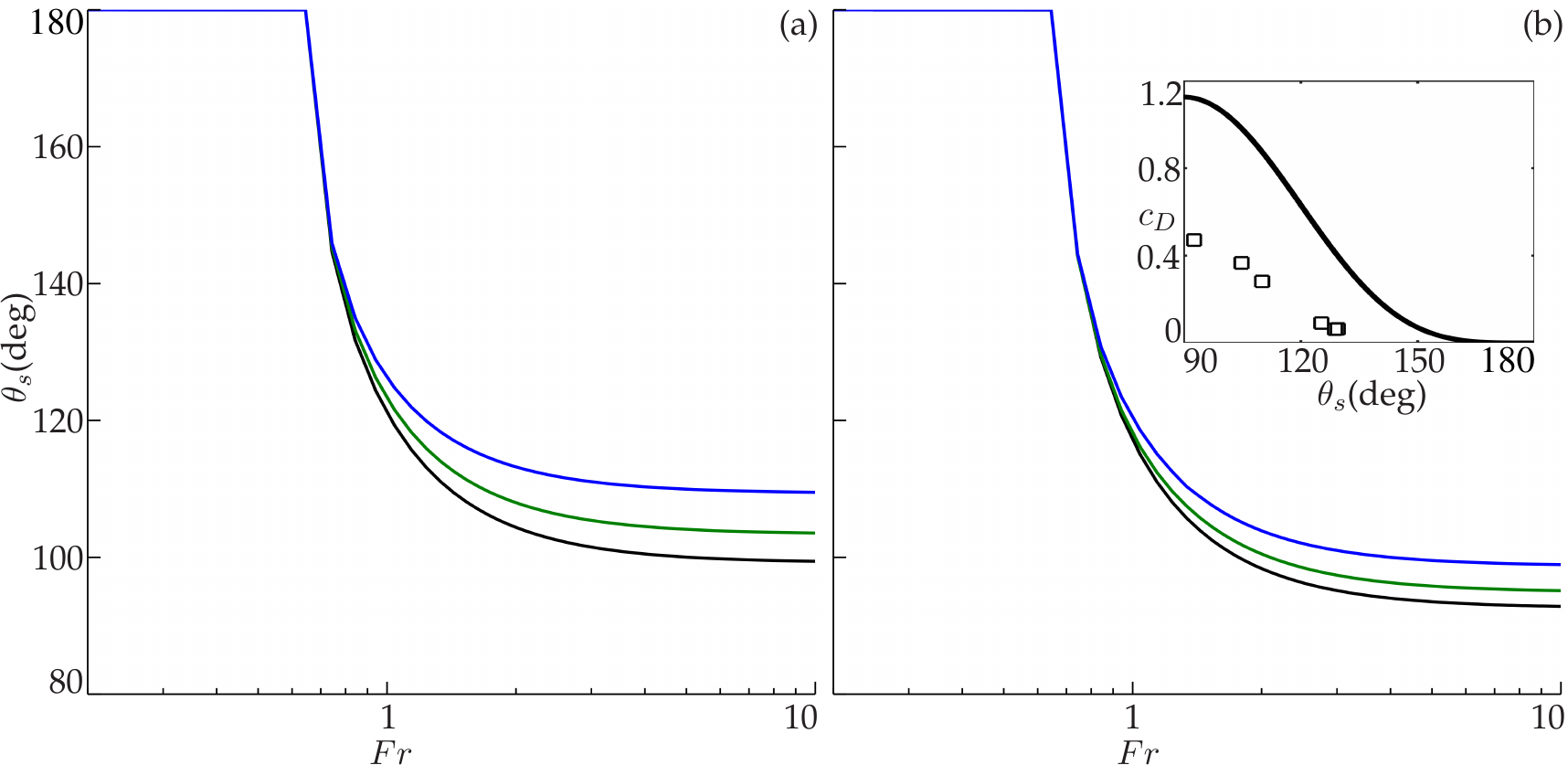}%
\caption{Separation angle, $\theta_s$, as a function of the Froude number, $Fr$, computed by integrating the boundary-layer equations~\eqref{cont_liquid_eq}--\eqref{energy_eq_interface}. (a) $T_\infty=50^{\circ}$C and wall temperatures of \com{$T_s=400^{\circ}$C} (blue line), \com{$T_s=500^{\circ}$C} (green line), and \com{$T_s=600^{\circ}$C} (dark gray line). (b) $T_\infty=75^{\circ}$C and wall temperatures of $T_s=400^{\circ}$C (blue line), $T_s=500^{\circ}$C (green line), and \com{$T_s=600^{\circ}$C} (dark gray line). \com{The inset of panel (b) represents the drag coefficient as a function of the separation angle, $C_D(\theta_s)$, where symbols are experimental values extracted from~\cite{Vakarelski2011}, while the solid line represents the function $9/8\sin^4 \theta_s$, obtained by integrating the pressure distribution of the outer potential flow, and assuming that the pressure remains constant downstream of the separation point.}\label{fr_vs_theta}}
\end{center}
\end{figure}

The configuration considered herein has been extensively employed in the past to study both natural- and forced-convection film-boiling regimes. However, most of these studies were mainly focused on the measurement of the heat transfer coefficient, and only a few authors paid attention to the separation phenomenon with the level of detail needed for a reliable quantitative comparison with the results presented herein. In particular, \citet{Zvirin1990} performed experiments with water under different conditions, observing two different regimes, either the formation of a vapour wake that separates from the sphere near its equator, or the formation or microbubbles. \citet{Liu1996} also performed experiments with water where they observed that, under certain conditions, the vapour layer remained attached along the whole sphere surface. More recently, \citet{Vakarelski2014} have carried out experiments with solid spheres falling in water. Since in all these studies the values of the Froude number are moderately large, we will devote the remainder of this section to discuss the effect of buoyancy on the separation of the flow.

To determine the numerical angle of separation with a unique criterion that covers all the cases explored, we decided to use the angle $\theta_s$ at which the vapour shear stress vanishes at the wall. Indeed, it was observed that, depending on the values of the vapour-to-liquid density and viscosity ratios, the relative importance of the convective acceleration compared with the viscous force in the vapour boundary layer can be small enough to allow the numerical computation of a substantial region of the recirculating vapour bubble. In the latter cases, the vapour stream behaves as a lubrication layer with small convective inertia. However, in other cases, the convective and viscous terms become of the same order near $\theta_s$, leading to a numerical singularity that prevents the computation of the vapour recirculation bubble. Nevertheless, it should be kept in mind that the actual separation angle, $\theta_f$, is only slightly larger than $\theta_s$ due to the explosive increase of the interfacial slope past $\theta_s$, which is therefore an appropriate measure of the separation angle in a first approximation.

When the liquid flow is opposite to gravity, as assumed in the present work, buoyancy forces tend to accelerate the vapour stream along the entire sphere. More specifically, equation~\eqref{azimth_mom_vapour_eq} indicates that the buoyancy force acting on the vapour stream is $O[(\rho/\rho_v) Fr^{-2}]$, while the pressure force is $O(\rho/\rho_v)$, indicating that $Fr^{-2}$ measures the relative importance of buoyancy forces compared to pressure forces. These two forces contribute to accelerate the vapour stream in the region \com{$0\leq \theta \leq 90^{\circ}$}, but they have opposite effects in the rearward half of the sphere. Indeed, the competition between the buoyancy force and the adverse pressure gradient is expected to play a significant role in delaying separation which, as discussed above, is mostly controlled by the onset of vapour recirculation. The function $\theta_s(Fr)$ plotted in figure~\ref{fr_vs_theta} reveals that, for $Fr\lesssim 1$, the buoyancy force overcomes the deceleration caused by the adverse pressure gradient, and avoids the formation of a vapour recirculating bubble, except in a very small region surrounding the rear stagnation point, \com{$\theta=180^{\circ}$}. As the Froude number increases, separation takes place upstream of the rear stagnation point at an angle $\theta_s(Fr)$ that decreases with increasing $Fr$, until an asymptotic angle is reached corresponding to the limit of negligible buoyancy forces discussed in \S\ref{subsec:fr_inf}. \com{As previously mentioned, the experiments show a giant drag reduction effect associated with the delayed separation. To derive a minimal model for the dependence of the drag coefficient, $C_D$, on the separation angle, $\theta_s$, we assume that the pressure distribution along the sphere is given by the potential flow solution for $0 \leq \theta\leq \theta_s$, and that it remains constant and equal to its value at $\theta=\theta_s$ for $\theta_s\leq \theta\leq \pi$. The latter assumptions, which are admittedly very strong, yield the result $C_D=9/8\sin^4\theta_s$, represented with a solid line in the inset of figure~\ref{fr_vs_theta} together with several experimental results extracted from~\cite{Vakarelski2011} (symbols). Although the model captures the strong dependence of the drag reduction on the separation angle, it overestimates the experimental drag coefficient by a factor of almost 3. The latter discrepancy is probably due to the fact that the pressure in the recirculating vapour region is larger than the potential value assumed in the model.}

To compare our model predictions with experimental measurements of the separation angle, we have performed numerical integrations using the working conditions of~\citet{Zvirin1990},~\citet{Liu1996} and~\citet{Vakarelski2014}, obtaining the results shown in table~\ref{table_1}. In all these cases, the numerical onset of the vapour recirculating bubble occurs at an angle slightly larger than \com{$90^{\circ}$}, in reasonable quantitative agreement with the experiments of~\citet{Zvirin1990}. Indeed, these authors reported close-up photographs of the sphere showing a large vapour wake departing from an angle that we estimated by image analysis. A similar vapour cavity was also reported by~\cite{Vakarelski2017} following the impact of a sphere at a temperature above the Leidenfrost point into water at 95$^\circ$C. \cite{Liu1996} also observed a long vapour wake departing from the equator when water was close to the saturation temperature. All these observations of the detachment of the vapour cavity close to the equator of the sphere are consistent with the numerical results summarised in table~\ref{table_1}, where the onset of vapour recirculation takes place close to \com{$\theta \geq 90^{\circ}$}. Unfortunately, the rapid downstream growth of the vapour bubble, and the resulting non-slender, unsteady two-phase wake cannot be described using our boundary-layer formulation. An appropriate theoretical and numerical description of the such flow seems a formidable task that is out of the scope of the present contribution.

\begin{center}
\begin{table}
\centering
\begin{tabular}{ c c c c c c }
 Reference & \com{$T_s$} ($^{\circ}$C) & $T_\infty$ ($^{\circ}$C) & $Fr$ & $\theta_s$(deg) & $\theta_s^{\text{exp}}$(deg) \\
 \cite{Zvirin1990} & 640 & 97.5 & 5.8 & 92.1 & 92.5\\
 \cite{Zvirin1990} & 579 & 70 & 6.14 & 95.12 & 97.8\\
  \com{\cite{Zvirin1990}} & \com{695} & \com{61.5} & 
  \com{6.49} & \com{94.8} & \com{119.1}\\
 \cite{Liu1996} & 293 & 91.5 & 1.47 & 103.11 & --\\
 \cite{Liu1996} & 328 & 91.3 & 2.11 & 96.68 & --\\
 \cite{Liu1996} & 426 & 90.9 & 4.56 & 91.65 & --\\
  \cite{Vakarelski2014} & 300 & 95 & 12.14 & 90.79 & --\\
 \cite{Vakarelski2014} & 500 & 85 & 11.18 & 91.15 & --
\end{tabular}
\caption{Summary of the experiments performed with water in the inverse Leidenfrost regime around a sphere at high Reynolds numbers. The value of $\theta_s$ was obtained from the numerical integration of~\eqref{cont_liquid_eq}--\eqref{energy_eq_interface}. The values of $\theta_s^{\text{exp}}$ were estimated from photographs reported by~\citet{Zvirin1990}.}
\label{table_1}
\end{table}
\end{center}

\section{Conclusions}

The significant drag reduction effect in the flow of cold liquids around solid spheres heated above the Leidenfrost point has been linked to a delay of the separation angle with respect to the case without a lubricating vapour layer surrounding the sphere, causing a decrease in the form drag~\citep{Vakarelski2011,Vakarelski2014}. In the present work, with the aim at explaining the experimental findings of~\citet{Vakarelski2011,Vakarelski2014,Vakarelski2016}, we have studied the flow in the high-Reynolds-number regime making use of boundary-layer theory. Our results indicate that the tentative explanation provided by~\cite{Vakarelski2011}, whereby the presence of the vapour layer effectively transforms the no-slip boundary condition at the wall into a stress-free boundary condition, does not explain the observed separation behaviour. Indeed, the hypothesis of an effectively stress-free interface reduces the problem to the flow around a spherical bubble at high Reynolds numbers, in which there is no boundary-layer separation except in a very small region close to the rear stagnation point~\citep{Moore1963}. Our findings, as pinpointed in \cite{Vakarelski2016,Berry2017}, reveal the central dynamical role played by the vapour layer in the rapid growth of the interface and the formation of a vapour wake. As a consequence, the use of an effective slip length~\cite{Vakarelski2016,Berry2017} is not a good approximation to the actual flow, since the slip length does not remain constant along the sphere. Moreover, under realistic parameter combinations, we have shown that the effective slip length may become locally negative, or even singular.

In contrast with the prevailing explanations, we have revealed that the separation of the flow in the inverse Leidenfrost regime is \com{profoundly affected} by the thin vapour layer surrounding the sphere. In particular, we have identified two key mechanisms that hinder the downstream transport of vapour, and force the vapour layer to grow explosively past a certain angle. Indeed, for \com{$\theta >90^{\circ}$}, the area per unit streamwise length is reduced as the rear stagnation point is approached. In addition, the adverse pressure gradient decelerates the vapour stream, eventually leading to the appearance of a recirculation bubble that forces the liquid-vapour interface to move away from the wall due to vapour mass conservation. Although condensation also takes place at the interface when the vapour layer becomes thick enough, the condensation rate is not enough to overcome both the geometric blocking effect and the adverse pressure gradient, eventually leading to the separation of the flow from the wall. For sufficiently small values of the Froude number, buoyancy forces can overcome these effects, and avoid the formation of the recirculation bubble, whereas for moderately large values of the Froude number, typical of most experimental conditions, buoyancy forces are only able to delay the onset of recirculation up to a certain angle \com{$\theta_s(Fr)<180^{\circ}$}. Our numerical results compare favourably with the experiments of~\citet{Zvirin1990},~\citet{Liu1996} and~\cite{Vakarelski2011}. Nevertheless, new experiments are needed for a more precise measurement of the separation angle, and for a systematic exploration of buoyancy effects.

\com{We would finally like to point out that future mathematical models aimed at describing the inverse Leidenfrost regime with a higher fidelity than the one reported herein, should contemplate the fact that the downstream evolution of the vapour layer must be obtained as part of the solution, coupling its governing equations with those of the liquid. In the present work, the vapour flow is described with the full boundary-layer equations, including the vapour convective inertia, what requires to solve the problem numerically. Inertial effects in the vapour stream become especially important in the region of backflow, where the vapour layer thickness experiences a large increase. However, in cases where separation does not take place until the rear stagnation point is reached, like those shown in figure~\ref{fr_vs_theta} for small enough values of the Froude number, it might well be the case that a simplified description of the vapour layer, in which convective inertia is neglected, provides a good leading-order description. Note that, under the latter approximation, the vapour flow would be described as a linear lubrication layer, which can be explicitly solved as the addition of a Couette flow induced by the polar velocity at the interface, and a Hagen-Poiseuille flow induced by the pressure gradient of the outer potential flow~\citep{Liu1996}. This simplification would avoid the need for a numerical calculation of the vapour flow, and only the liquid stream would have to be computed numerically. However, it must be emphasized that the latter simplified description is probably not valid to account for separation.}

\begin{acknowledgments}
J.A. thanks the Govern de les Illes Balears for financial support through the Vicen\c{c} Mut subprogram partially financed by the European Social Fund and the Spanish MCINN for support through grant CTM-2017-83774-D. A.S. thanks the Spanish MINECO for financial support through projects DPI2015-71901-REDT and DPI2017-88201-C3-3-R, partly financed through European funds. The authors acknowledge helpful comments and suggestions by Professor Norman Riley and Professor Eduardo Fern\'andez-Tarrazo. J.A. acknowledges insightful discussions with Dr. Idan Tuval. The detailed and constructive comments of an anonymous reviewer, which have contributed to improve the present work, are gratefully acknowledged.
\end{acknowledgments}

\vspace{5mm}
\textbf{Declaration of Interests.} The authors report no conflict of interest.

\end{document}